# Acousto-optic modulation based on an AlScN microring resonator for microwave-to-optical conversion

**Author:** Kewei Bian[1], Yushuai Liu[2,3,4], Weilin Rong[1], Yuan Dong[1], Qize Zhong[1], Yang Qiu[1], Xingyan Zhao[1], Tao Wu[2,3,4], Shaonan Zheng[1]* and Ting Hu[1]*

[1] *School of Microelectronics, Shanghai University, Shanghai 201800, China*
[2] *School of Information Science and Technology, ShanghaiTech University, Shanghai, 201210, China*
[3] *Shanghai Institute of Microsystem and Information Technology, Chinese Academy of Sciences, Shanghai, 200050, China*
[4] *University of Chinese Academy of Sciences, Beijing, 100049, China*
* snzheng@shu.edu.cn *hu-t@shu.edu.cn;

**Abstract:** Acoustic-optic (AO) modulation is critical for microwave and optical signal processing, computing and networking. Challenges remain to integrate AO devices on-chip using fabrication process compatible with complementary metal-oxide-semiconductor (CMOS) technology. This work presents the demonstration of an AO modulator exploiting a microring resonator (MRR) based on thin-film aluminum scandium nitride (AlScN) photonic platform. Leveraging the high piezoelectric properties of AlScN, an MRR is employed with interdigital transducer (IDT) inside to couple microwave signals into acoustic resonant modes, enabling efficient by-directional optical modulation in the MRR. The fabricated MRR exhibits an optical loaded quality factor ($Q$) of $1.8 \times 10^4$ at the optical L-band for the $TE_{00}$ mode. A low effective half-wave voltage ($V_\pi$) of 1.21 V is achieved, corresponding to a $V_\pi L$ of 0.0242 V·cm, along with an optomechanical single-photon coupling strength $g_0$ of 0.43 kHz between the 2.11 GHz acoustic mode and the $TE_{00}$ optical mode. The device shows potential for applications in microwave photonics.

## 1. Introduction

Acousto-optic (AO) modulation relies on the interaction between phonons and photons, where phonons manipulate photons by altering the refractive index of the medium.[1-6] Leveraging this principle, AO devices find wide applications in optical frequency shifting,[7-9] light deflection,[1,10] optical isolation,[2,3,5,11] and microwave-to-optical conversion.[12-15] Notably, conversion between microwave and optical information is a critical technology for classical hybrid signal processing and interfacing superconducting quantum computers with quantum networks.[16] AO devices can couple microwaves into acoustic resonant modes by using electromechanical transducers, achieving the transfer of information from microwave to optical domains through the acousto-optic effect. Optomechanical systems, based on optical waveguides in microwave cavities,[16,17] membranes in free-space Fabry-Pérot cavities,[18-20] and nanoscale piezoelectric optomechanical crystals,[21-24] have demonstrated significant potential in microwave-to-optical conversion and microwave signal processing.

To date, most of the integrated AO devices have been realized using piezoelectric materials, such as lead zirconate titanate (PZT),[25] gallium nitride (GaN),[26,27] thin-film lithium niobate (TFLN),[28-30] and aluminum nitride (AlN).[31] Structures adopted in piezoelectric tuning mainly include waveguide, MRR, and Mach-Zehnder interferometer (MZI). Compared to waveguide and MZI-based counterparts, MRR-based AOM is more compact and exhibits higher modulation efficiency due to the resonant characteristics, while the optical bandwidth is smaller.

The fabrication processes of PZT and GaN are not compatible with the complementary metal-oxide-semiconductor (CMOS) technology. AlN films have the potential to realize AO modulators with low cost as their fabrication process is compatible with CMOS technology.[32,33] The piezoelectric properties of AlN films can be significantly improved by doping with scandium (Sc) elements, which has been widely demonstrated in the development of piezoelectric microelectromechanical systems (MEMS).[34-36] Furthermore, recent studies indicate that AlScN exhibits a wide electronic band gap comparable to that of AlN, resulting a broad transparent window from ultraviolet to mid-infrared.[38,39] High second-order optical nonlinearity and electro-optic effect have been experimentally demonstrated.[39,40]

Compared to AlScN, TFLN presents much higher electro-optic coefficient, much lower optical loss, and comparable piezoelectricity, which has attracted extensive attention in research and industry. Nevertheless, LN on insulator (LNOI) wafer with low cost and large size up to 8 inch is not yet commercially available. In contrast, AlScN can be grown directly through physical vapor deposition (PVD) process on a wafer up to 8 inch and be easily heterogeneously integrated with common materials such as silicon (Si), silicon nitride, germanium, etc. on the same Si wafer without bonding process. Hence, AlScN shows fabrication cost-effectiveness and versatility. Therefore, AlScN has emerged as a promising platform, of which the potential needs to be further unveiled for integrated AO devices.

This work demonstrated an AO modulator using a microring resonator (MRR) based on a thin-film AlScN photonic platform for the first time. By taking advantage of the high piezoelectricity properties and CMOS-compatibility of AlScN, microwave-to-optical conversion is achieved. Using the interdigital transducer (IDT), the device couples microwave input to the acoustic resonance mode, enabling effective optical modulation through the MRR. Specifically, the MRR with an optical $Q$ factor of $1.8 \times 10^4$ is modulated using an acoustic resonator with a $Q$ factor of 650 at a frequency of 2.11 GHz, achieving an effective half-wave voltage ($V_\pi$) of 1.21 V and an optomechanical single-photon coupling strength of 0.43 kHz.

## 2. Results

The schematic diagram of the AO modulator is shown in Figure 1a. An AlScN racetrack MRR is employed. The IDT and the MRR are integrated on the thin-film AlScN with a scandium concentration of 9.6% (Sc: Al =9.6%: 90.4%, atomic percentage). The IDT is placed inside the MRR, driven by the RF signal to excite a bidirectional surface acoustic wave (SAW) propagating in the $x$ direction. Two acoustical reflectors consisting of 100 pairs of aluminum metal electrodes each are placed at 5 times of acoustic wavelength away from the MRR to reflect the acoustic wave, forming a standing wave in the waveguide area, thereby improving the AO modulation efficiency of the device. The schematic diagram of the modulation principle is shown in Figure 1b. The probe light with angular frequency of $\omega_p$ is input from the optical input port.

Under the modulation of the SAW at frequency of $\Omega_m$, the probe light is scattered to frequency-detuned waves at $\omega_p \pm \Omega_m$ in the same optical transverse electric mode $TE_{00}$.

Figure 1c illustrated the cross section of the ridge waveguide area. The optical waveguides are fabricated by partially etching a 400-nm-thick AlScN film with an etching depth of 220 nm and the top width of the AlScN waveguide is 1 μm. The numerical simulation result of the normalized field distribution of the optical mode shows that the fundamental transverse electric ($TE_{00}$) optical mode is supported. The IDT consists of 100 pairs of aluminum electrodes with a period $\Lambda$ of 2.2 μm to excite acoustic wave with wavelength of 2.2 μm. The electrode width is 550 nm, and the aperture is 200 μm. The numerical simulation results of the normalized displacement components in the x-direction ($u$) and z-direction ($v$) of the SAW mode with a frequency of 2.1 GHz are shown in Figure 1d.

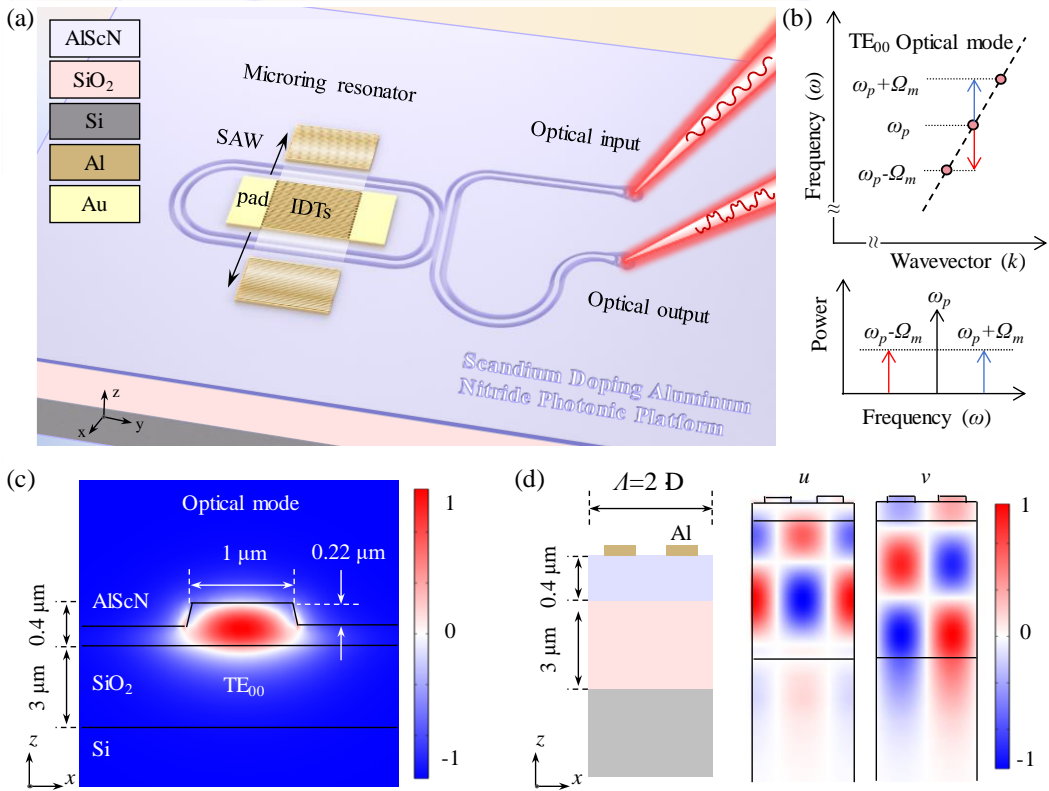

**Figure 1.** (a) The schematic diagram of the MRR AO modulator based on the thin-film AlScN platform. (b) The schematic diagram of the modulation principle. (c) Electric field of the $TE_{00}$ optical mode. (d) Numerical simulation results of normalized displacement components in the x-direction ($u$) and z-direction ($v$) of the SAW mode in single-cycle IDT region.

The fabrication process flow for the AlScN AO modulator is shown in Figure 2a. A 3-μm-thick $SiO_2$ layer is grown by thermal oxidation on an 8-inch Si wafer. Next, 400-nm-thick AlScN is deposited on the wafer by magnetron sputtering from a pre-alloyed target (Sc: Al=9.6%: 90.4%, atomic percentage). Detailed AlScN film characterization can be found in our previous work.[41,42] Electrical contact pads for IDT and alignment marks for subsequent electron beam lithography (EBL) alignment are patterned in a 100-nm-thick gold (Au) layer on the top of AlScN using a lift-off process.

Then a 450-nm-thick SiO$_2$ layer is deposited as hard mask for AlScN etching. After EBL, the optical ridge waveguides with a designed width (*W*) are fabricated by dry etching of the AlScN layer. Finally, a lift-off process is employed again to form the 100-nm-thick aluminum IDT on top of the AlScN layer, where the IDT consists of equidistant fingers with a pitch of 1.1 μm. This AlScN photonic platform allows the fabrication of optical waveguides and electromechanical transducers within the same material layer.

Figure 2b shows the scanning electron microscope (SEM) image of the Al IDT. The period is measured as 2.15 μm, which is smaller than the target value of 2.2 μm. Figure 2c shows the SEM image of the AlScN ridge waveguide. The cross section of the AlScN ridge waveguide SEM image is shown in Figure 2d. The sidewall angle is 69.8° (left) and 62.5° (right). The top width is 960.5 nm and the etch depth is 221.1 nm. The deviations from the designed value are mainly due to the fabrication error.

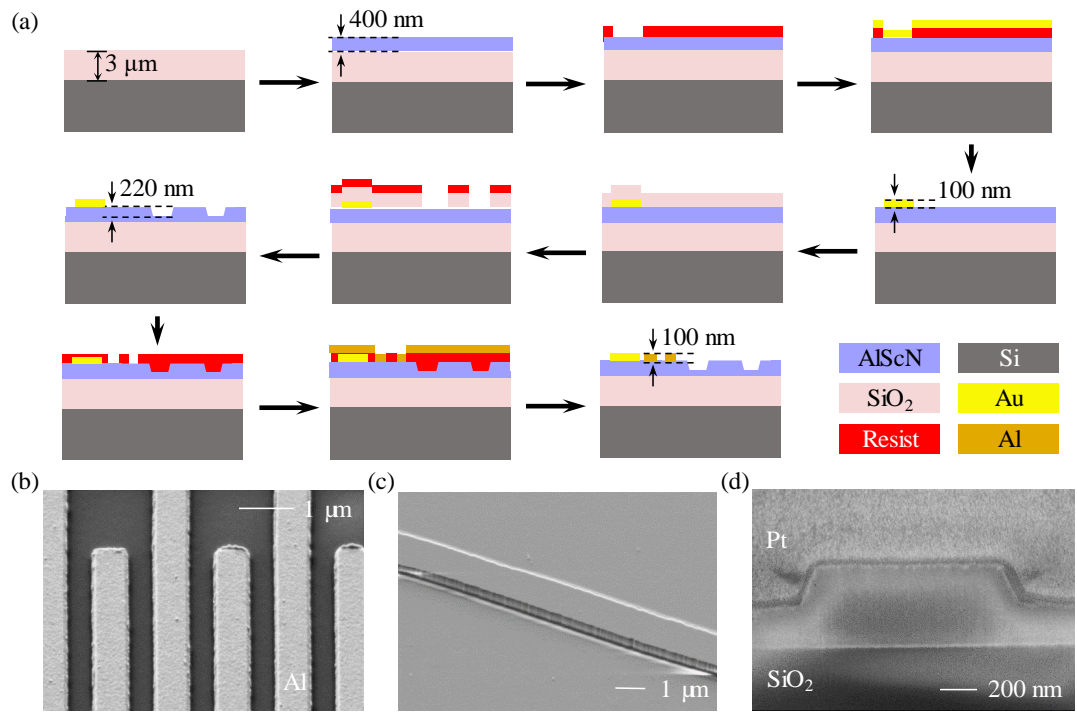

**Figure 2.** (a) Fabrication process flow of the AlScN AO modulator. SEM images of the fabricated (b) Al IDT, (c) AlScN ridge waveguide, and (d) cross section of the AlScN ridge waveguide.

Figure 3a-c shows the optical microscopic images of the fabricated device, the magnified image of the optical direction coupler and the AO interaction area, respectively. The length of the straight arm of the racetrack MRR is 420 μm, while the length of the direction coupler coupling region is 40 μm. The short and long bending radii of the racetrack MRR are 100 μm and 120 μm, respectively. The optical transmission spectrum of the MRR is extracted by scanning at a wavelength near 1570 nm using a tunable laser. The transmission spectrum results of the TE$_{00}$ optical mode are shown in Figure 3d. Figure 3e presents a normalized single resonance peak of MRR at 1574.1 nm. By performing a Lorentz fitting on the transmission spectrum, the *Q* factor of the device is calculated to be $1.8 \times 10^4$. The average propagation loss of the

MRR waveguide is calculated to be 10.8 dB/cm including the bending loss,[43] which is mainly caused by the rough sidewalls of the waveguide. Since there is no upper cladding structure the scattering of particles in the air cladding above the waveguide would also contribute to an increase in optical loss. By further optimizing the film quality and etching process, it is expected to achieve higher $Q$ factor in the MRR to significantly improve the modulation efficiency of the device.

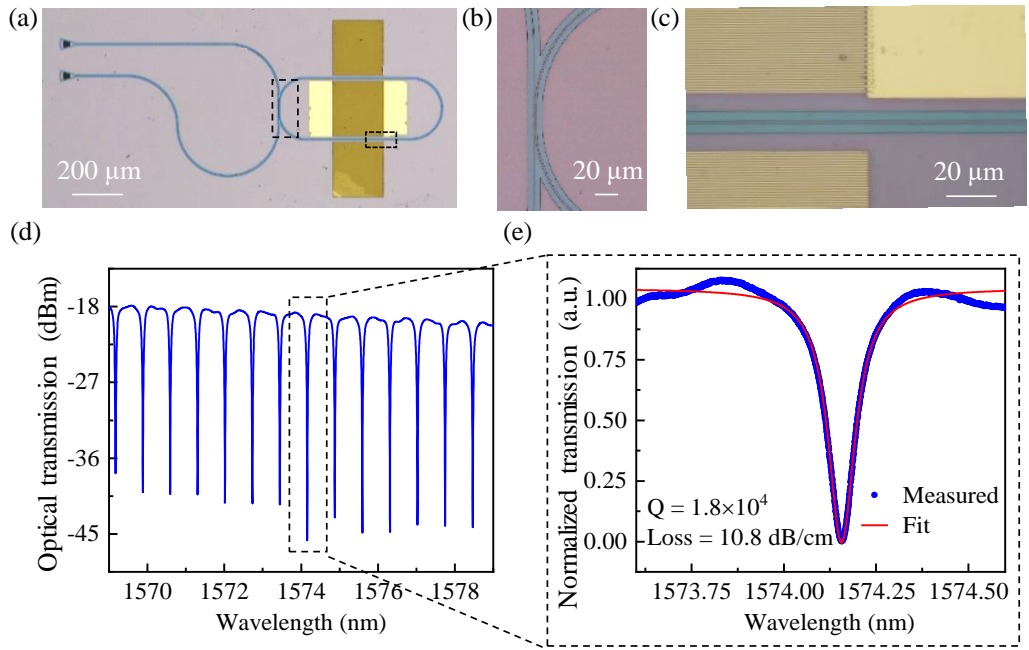

**Figure 3.** (a) Microscopic image of the MRR AO modulator. The magnified images show (b) the optical direction coupler and (c) the AO interaction area. (d) Measured transmission spectrum of the MRR for $TE_{00}$ mode. (e) Lorentz fitting (red curve) of the resonance dip at 1574.155 nm, which corresponds to a loaded $Q$ factor of $1.8\times10^4$.

The AO modulation of the device is characterized by experimentally measuring the acousto-optic $S_{21}$ spectrum. The experimental setup is shown in Figure 4. Probe light from the tunable laser (TL, Keysight 81634B) propagates through a single-mode fiber, with its polarization direction (transverse electric, TE) adjusted using a fiber polarization controller. The optical signal couples into and out of the AO device through grating couplers, which have an insertion loss of 9.02 dB/facet. Finally, the output optical signal is detected by a photodetector (PD, Thorlabs RXM10AF). Port 1 of the vector network analyzer (VNA, Keysight ENA E5071C) is connected to the IDT to apply RF signal. Port 2 of the VNA is connected to the PD to acquire the PD signal. By adjusting the wavelength of the incident light, the MRR gradually detunes from the resonant state, and a progressively increasing $S_{21}$ value can be observed on the VNA. The pump laser wavelength is 1571.13 nm where the $S_{21}$ reaches a maximal value.

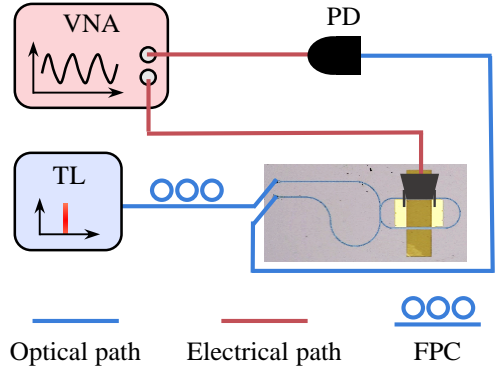

**Figure 4.** Schematic diagram of the device measurement system. TL: tunable laser; VNA: vector network analyzer; FPC: fiber polarization controller; PD: photodetector.

The RF to acoustic conversion efficiency of the IDT can be evaluated by measuring its reflection coefficient ($S_{11}$). As shown in Figure 5, the IDT exhibits several resonance peaks within the frequency range of 1-3 GHz. At the peak of 2.11 GHz, the acoustic resonance has a $Q$ factor of 650, corresponding to a linewidth of 3.12 MHz. $\Delta|S_{11}|^2$ is calculated to be 41%, which represents the ratio of the RF power loaded on the IDT to the actual input RF power.[44] When the RF power loaded on the IDT is 21 dBm, the opto-acoustic $S_{21}$ spectrum has three significant peaks at the frequencies of 1.82 GHz, 2.11 GHz and 2.36 GHz, which agrees well with the peaks of the $S_{11}$ spectrum, indicating that the AO modulation of the device is strong at these frequencies. At the frequency of 1.82 GHz, $\Delta|S_{11}|^2$ is less than 15%. By optimizing the electrodes of IDT to meet the impedance matching conditions at this frequency, the $S_{21}$ value can be further improved.

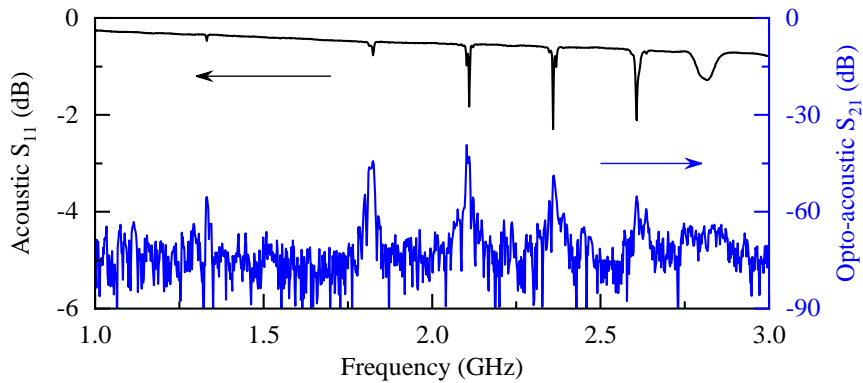

**Figure 5.** Acoustic $S_{11}$ spectrum (black line) of the IDT and opto-acoustic $S_{21}$ spectrum (blue line) of the MRR AO modulator.

In order to quantify the AO modulation performance of the device, the effective half-wave voltage $V_\pi$ of the proposed AO modulator was extracted from experimental measurements of the $S_{21}$ spectrum,[29]

$$V_\pi = \frac{\pi R_{PD} I_{rec}}{|S_{21}|}, \qquad (1)$$

where $R_{PD}$ is the sensitivity of the optical receiver and $I_{rec}$ is half of the maximum DC

optical power. The experimental results show that $S_{21}$ is $-39.3$ dB at 2.11 GHz. The $R_{PD}$ of the PD is 450 V/W, and the $I_{rec}$ is 0.0093 mW. Hence the calculated effective half-wave voltage $V_\pi = 1.21$ V, and the product of voltage and modulation length $V_\pi \cdot L$ is 0.0242 V·cm. It is worth noting that the effective $V_\pi$ calculated here is different from the AO modulator based on Mach-Zehnder interferometer, because in the MRR, when the light transmittance changes from the maximum value to the minimum value, the phase difference does not change by $\pi$.[16,29]

Next, we determine the overall AO single-photon coupling strength $g_0$. For weak microwave inputs, the relation between the $S_{21}$ and $g_0$ is given by:[16]

$$S_{21} = \frac{8g_0^2 \gamma_e k_e^2 R_{PD}^2 I_{rec}^2}{\hbar \gamma^2 k^2 \left(\Omega_m^2 + \frac{k^2}{4}\right) \Omega_m R_{load}}, \quad (2)$$

where $\kappa$ ($\gamma$) and $\kappa_e$ ($\gamma_e$) are the total loss and external coupling rate of the optical (acoustic) mode, respectively. The parameter $\Omega_m$ is the frequency of the acoustic mode, and $R_{load} = 50\ \Omega$ is the impedance of the input microwave source. Equation 2 is derived from the equation of motion for the dynamics of the acousto-optic cavity. We estimate the acousto-optic single photon coupling strength to be $g_0 \sim 0.43$ kHz between the 2.11 GHz acoustic mode and the $TE_{00}$ optical mode. The parameters used for extraction of $g_0$ is shown in Table 1.

**Table 1. Parameters Used for Extraction of $g_0$**

| $k/2\pi$ | $k_e/2\pi$ | $\gamma/2\pi$ | $\gamma_e/2\pi$ | $\Omega_m/2\pi$ | $I_{rec}$ | $R_{pd}$ | $R_{load}$ | $S_{21}$ |
|---|---|---|---|---|---|---|---|---|
| 5.445 GHz | 4.845 GHz | 3.3 MHz | 3.12 MHz | 2.11 GHz | 0.0093 mW | 450 V/W | 50 Ω | −39.3 dB |

The AO device in this work is integrated on a silicon substrate and fabricated using CMOS-compatible technology without the need for suspended structures. The piezoelectric properties of the AlScN film are constrained by the limited concentration of Sc. By using AlScN with higher Sc concentrations, such as $Al_{0.57}Sc_{0.43}N$, which exhibits a piezoelectric coefficient $d_{33}$ of 27.6 pC·N$^{-1}$,[40] devices are expected to achieve more efficient electromechanical conversion.[32] However, at higher Sc concentrations, film structure becomes worse[45,46] and the etching process becomes more challenging, making it harder to control the roughness of the waveguide sidewalls. Hence, increased optical loss will be induced. Exploiting AlN as a seed layer before depositing AlScN with higher Sc concentrations will help improve film quality. Growing a layer of high-speed sound material on the wafer before deposition of the AlScN film will enhance the electromechanical coupling coefficient of the devices.

## 3. Conclusion

In summary, we demonstrate AO modulation using an MRR on thin-film AlScN photonic platform. The fabricated MRR demonstrates a $Q$ factor of $1.8 \times 10^4$ at optical L-band for the $TE_{00}$ mode. The IDT is fabricated on the AlScN surface to take full advantage of its piezoelectric properties. An effective $V_\pi$ of only 1.21 V which corresponds to a $V_\pi L$ of 0.0242 V·cm is observed. The acousto-optic racetrack MRR cavity exhibits an optomechanical single-photon coupling strength of 0.43 kHz. The presented work promises the potential of the AlScN photonics platform for microwave photonics applications.

**Funding.** National Natural Science Foundation of China (Grant No. 62205193, No. U23A20356, and No. 62204149), Natural Science Foundation of Shanghai (Grant No. 23ZR1442400), and Jiangsu Provincial Key Research and Development program (BE2023048).

**Notes.** The authors declare no competing financial interest.

**Data availability.** Data underlying the results presented in this paper are not publicly available at this time but maybe obtained from the authors upon reasonable request.

**Acknowledgments.** This work was supported by Shanghai Collaborative Innovation Center of Intelligent Sensing Chip Technology, and Shanghai Key Laboratory of Chips and Systems for Intelligent Connected Vehicle.